# Magnetic Relaxation in Bismuth Ferrite Micro-Cubes

B. Andrzejewski, K. Chybczyńska, K. Pogorzelec-Glaser, B. Hilczer, T. Toliński
Institute of Molecular Physics
Polish Academy of Sciences
Smoluchowskiego 17, PL-60179 Poznań, Poland
bartlomiej.andrzejewski@ifmpan.poznan.pl

B. Łęska, R. Pankiewicz
Faculty of Chemistry
Adam Mickiewicz University
Umułtowska 89b, PL-61614 Poznań, Poland

P. Cieluch
Research Centre of Quarantine, Invasive and Genetically Modified Organisms
Institute of Plant Protection – National Research Institute
Węgorka 20, PL-60318 Poznań, Poland

*Abstract*— **The process of magnetic relaxation was studied in bismuth ferrite BiFeO3 multiferroic micro-cubes obtained by means of microwave assisted Pechini process. Two different mechanisms of relaxation were found. The first one is a rapid magnetic relaxation driven by the domain reorientations and/or pinning and motion of domain walls. This mechanism is also responsible for the irreversible properties at low temperatures. The power-law decay of the magnetic moment confirms that this relaxation takes place in the system of weakly interacting ferromagnetic or superferromagnetic domains. The second mechanism is a longterm weak magnetic relaxation due to spin glass-phase.**
*Keywords-component; magnetic relaxation, domain walls, spin-glass, bismuth ferrite*

I. INTRODUCTION

Bismuth ferrite $BiFeO_3$ (BFO) belongs to a group of materials called magnetoelectric (ME) multiferroics, that exhibit charge and magnetic ordering with some mutual coupling between them [1-4]. ME multiferroics recently have attracted the attention of numerous groups of researchers because of their very interesting and rich physical properties and prospective technological potential [5]. BFO compound is a rhombohedrally distorted perovskite with space group $R3c$ at room temperature. The ferroelectric properties appear in BFO below Curie ferroelectric temperature $T_C$=1100 K due to charge ordering caused by the ordering of lone electron pairs of $Bi^{3+}$ ions. The magnetic properties together with weak ferromagnetic (FM) moment results from the complex ordering of $Fe^{3+}$ spins and they appear below the Néel temperature $T_N$=643 K. The magnetic ordering in BFO is an antiferromagnetic state (AFM) exhibiting G-type structure and superimposed long range incommensurate cycloidal modulation with the period $\lambda$=62 nm [3, 6]. The spin cycloid propagates along three equivalent crystallographic directions [1,-1,0], [1,0,-1] and [0,-1,1] (pseudocubic notation). The spins in the cycloid rotate in the plane determined by the direction of cycloid propagation and the [1,1,1] direction of spontaneous electric polarization. ME type of Dzyaloshinskii-Moriya interaction induces small canting of the spins out of the rotation plane, which leads to a local ferromagnetic ordering. This local FM ordering in the form of weak ferromagnetic domains was indeed recently found experimentally in BFO by means of neutron diffraction [7]. The mean size of the domains is 30 nm *i.e.* a half of modulation period of the spin cycloid.

A system containing magnetic domains or clusters like BFO can exhibit variety of magnetic orderings depending on the strength of interdomain interactions. For negligible energy of interaction, the system attains a superparamagnetic (SP) state. With increasing energy of interdomain coupling the system start to attain collective magnetic states; first a superspin glass (SSG) state for moderate interactions and next superferromagnetic (SFM) state if the coupling becomes strong enough [8]. Moreover, the properties of SFM or FM systems are modified by growth and/or reorientations of magnetic domains and also by pinning or motion of domain walls which makes their behavior very complex. Magnetic domain walls can be pinned at pinning centers like; local lattice strains, structural defects, grain boundaries, and in multiferroics even at ferroelectric domains due to flexomagnetic interactions [9]. This last phenomenon is another interesting example of the coupling between electric and magnetic properties in multiferroics. Domain walls are bowed between the pinning centers but they can move or jump to the another center due to magnetic interaction or thermal excitations. The motion of magnetic domain walls causes domain growth, domain wall reconformations and also relaxation of a magnetic moment.

Besides weak ferromagnetism and ferromagnetic domains found in BFO [7], a spin-glass (SG) transition at low temperatures has been also recently postulated by Singh et al. [10, 11]. However, as noted in [10, 11], it is very difficult to distinguish SG phase from SP one and from ferroics with domain wall pinning and motion or relaxors, because all these systems exhibit existence of Almeida de Thouless line (AT-line) [11], aging, rejuvenation, magnetic relaxation and sometimes also the memory effect.

The aim of this paper is to shed some light on the question which phenomenon brings a dominant effect on the magnetic properties of BFO: SG phase or relaxation due to FM domain growth, domain reorientations and wall pinning or motion. This problem is very essential because it is expected that the domain walls in multiferroics will be active elements in future device applications due to extremely short switching time and low consumption of energy instead of devices using ferroelectric or ferromagnetic domains [12].

## II. EXPERIMENTAL

### A. Sample Synthesis

BFO powder-like samples were obtained by means of microwave assisted hydrothermal Pechini process. An aqueous solution necessary for the synthesis was prepared by dissolving appropriate amounts of nitrates; $Bi(NO_3)_3 \cdot 5H_2O$, $Fe(NO_3)_3 \cdot 9H_2O$, sodium carbonate $Na_2CO_3$ and potassium hydroxide KOH in distilled water. This solution was transferred and sealed in PTFE reactors. Next, it was processed in CEM Mars-5 microwave oven at 200 $^0$C for about 30 minutes. During the reaction, the water vapour pressure inside the reactors was about $2 \cdot 10^6$ Pa. After the reaction was completed, the oven was cooled gradually to about 50 $^0$C. Than the as obtained, brown in coloration suspension was filtered off to collect the fine BFO powder. This powder was rinsed with water, dried in air and some samples were also air calcined at 500 $^0$C for 1 hour.

### B. Sample Characterization

The crystallographic structure of the samples and minority phase content was studied by means of x-ray diffraction (XRD) using an ISO DEBYEYE FLEX 3000 diffractometer equipped with a Co lamp ($\lambda$=0.17928 nm). Their morphology was examined by S3000N Hitachi Scanning Electron Microscope (SEM), whereas magnetometric measurements were performed using the Quantum Design Physical Property Measurement System (PPMS) fitted with a superconducting 9T magnet and with a Vibrating Sample Magnetometer (VSM probe). Before measurements of magnetic relaxation, the superconducting magnet was switched to driven mode for about 1 h to remove the remnant magnetic flux.

## III. RESULTS AND DISCUSSION

Fig. 1 presents the XRD pattern of as-obtained microwave-synthesized product. The solid line corresponds to the best Rietveld profile fit to the experimental data calculated by means of FULLPROF software. This analysis reveals a well crystallized BFO rhombohedral phase with *R3c* space group and a small content of $Bi_{25}FeO_{40}$ parasitic phase labeled by asterisk. The line below the XRD data shows the difference between the experimental data and the fit. The vertical sections indicate the positions of Bragg peaks. The amount of $Bi_{25}FeO_{40}$ parasitic phase evaluated with respect to the most intensive BFO peaks (110, 104) is about 2%. The low content of the parasitic phase is a result of a short time of microwave reaction which promotes the synthesis of major BFO phase. Also it was observed that the $Bi_{25}FeO_{40}$ undesired phase is usually formed mainly during an early stage of the reaction at a low temperature. For microwave synthesis, duration of this stage of reaction is very short because of even heating of the solution volume and high rate of temperature increase in the microwave oven.

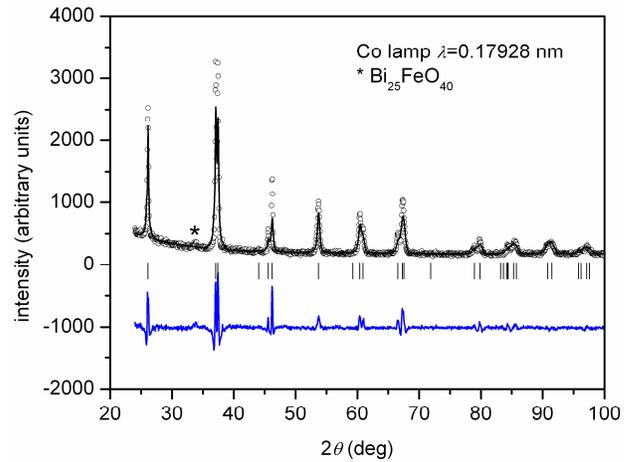

Figure 1.  XRD-pattern of BFO sample. The solid line is the best fit to the experimental data represented by open points. The solid line below is the difference between the data and the fit. The vertical sections represent positions of the Bragg peaks. The asterisk indicates the trace of $Bi_{25}FeO_{40}$ phase.

Fig. 2 presents a SEM micrograph illustrating the morphology of BFO sample. It turns out that the BFO grains are actually the agglomerates composed of almost regular micro-cubes of the mean size of about 1 μm, however the size distribution of the grains is rather wide. The magnetization loops of BFO sample $M(H)$ at three selected temperatures *i.e.* 10 K, 100 K and 300 K are shown in Fig. 3. Complete saturation of magnetization is never achieved even at the magnetic field 9 T, much stronger than that applied to record the data plotted in Fig. 3 (not presented here). At high temperatures 100 K and 300 K, the magnetization is reversible, however it exhibits pronounced hysteresis at 10 K. The shape of the hysteresis loops, its evolution with temperature and appearance of irreversibility in low temperatures resembles the behavior of SP, SSG phases reported by Shen et al. [13] or a weak ferromagnetic order caused by the presence of SFM phase. The presence of FM order was verified by means of Arrot construction [14] where the magnetization $M(H)$ data are plotted in the form of the square of magnetization $M^2$ versus dimensionless variable $H/M$. According to Arrot approach based on the Weiss molecular-field theory [15], the relation between magnetization $M(H)$ and internal magnetic field is as follows:

$$M^2 = \frac{3\mu M_0^3}{k_B T}\left(\frac{H}{M}\right) + 3\left(\frac{T_c}{T} - 1\right)M_0^2 \quad (1)$$

where $\mu$, $M_0$, $k_B = 1.38 \cdot 10^{-23}$ J/K and $T_c$ are the magnetic moment per atom, the spontaneous magnetization at zero temperature, Boltzmann's constant and the Curie temperature respectively.

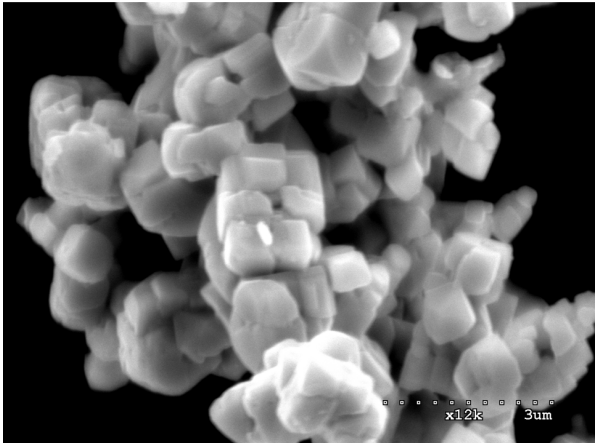

Figure 2. SEM micrograph of BFO sample.

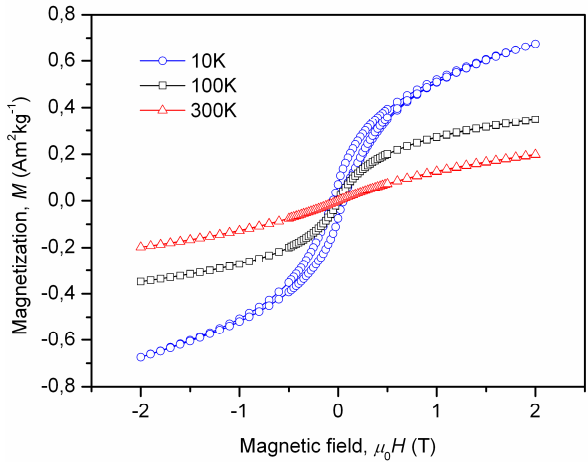

Figure 3. BFO magnetization $M(H)$ loops for the temperatures 10 K, 100 K and 300K.

If the data are plotted in the form of a square of the magnetization $M^2$ vs. $H/M$, they ordinate in the Arrot construction along the straight isotherms for each measuring temperature and can be fitted by a linear function. The slope of a linear fit is positive and increases with decreasing temperature. The intercept of the linear fits with the ordinate axis evolves from negative to positive when the magnetic ordering changes from paramagnetic to ferromagnetic. The intercept with the origin of the Arrot plot occurs for the isotherm corresponding to the Curie temperature. The Arrot construction for the BFO sample measured at a few temperatures selected from the range 2 K-300 K is shown in Fig. 4. The parts of isotherms recorded for high magnetic field are well fitted by linear functions and length of these linear sections of isotherms increases as the temperature decreases. However, for low magnetic fields the isotherms exhibit an apparent positive curvature, which can indicate a second-order-transition [16]. The ordinate intercept is always negative which means that there is no net ferromagnetic order. For the BFO sample studied, the intercept approaches the origin of the Arrot plot when temperature decreases, which indicates the increasing role of interdomain ferromagnetic interactions. However, these interactions are not strong enough to induce net FM ordering. The lack of evidence for FM ordering in the Arrot plot in Fig. 4 may be also explained in another way, in terms of a weak BFO ferromagnetism, which results from complex spin ordering. Without any distortion of the structure, BFO compound should be an ideal antiferromagnet with no net magnetic moment because all magnetic moments of $Fe^{3+}$ spins are compensated. Also the local magnetic moments due to spin canting in the spin cycloid should be balanced out by the moments of the neighboring cycloids. Then, the non-zero magnetic moment can originate only from the diminutive tilt of the moments in the adjacent (111) planes which induces weak canted ferromagnetism [17].

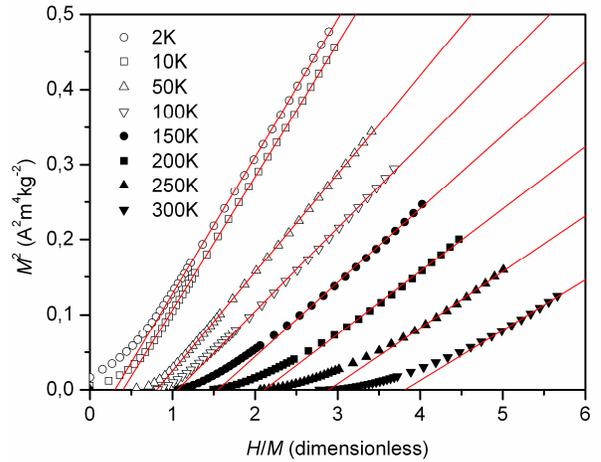

Figure 4. The Arrot plots for BFO sample at various selected temperatures.

The inverses of zero field cooling (ZFC) and field cooling (FC) susceptibilities $1/\chi(T)$ areshown in Fig. 5. Above about 70 K, the ZFC and FC inverses of susceptibility are identical and they deviate from the linear relation derived from the Curie-Weiss law: $1/\chi(T) = (T-\Theta)/C$ (where C is the Curie constant and $\Theta$ is the Weiss temperature). The nonlinear dependence of the inverse of susceptibility $1/\chi(T)$ usually is attributed to the formation of SP domains and to the increase in domain volume with decreasing temperature [13]. A different behavior is observed below 70 K, where the inverses of ZFC and FC susceptibilities start to bifurcate. In this temperature range the inverse of ZFC susceptibility increases with decreasing temperature, whereas the inverse of FC susceptibility is almost constant. This means that the value of FC susceptibility $\chi(T)$ also does not change at low temperatures, a feature that usually indicates the onset of interdomain interactions [18-23]. The nonzero interactions between domains lead to the appearance of a SSG or a SFM phase, if the interactions are strong enough.

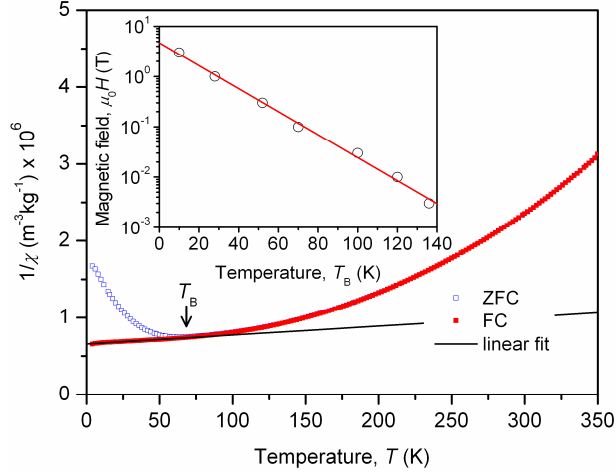

Figure 5. The inverse of ZFC and FC susceptibilities $1/\chi(T)$ (main panel) and the dependence of $T_B(H)$ temperature vs. field (inset). The solid line is the linear fit to the FC data at low temperature range.

The bifurcation between ZFC and FC susceptibilities, which appears at $T_B$ is usually explained in terms of SP domain blocking. The blocking temperature depends on the domain volume $V$ and the anisotropy constant K and can be evaluated from the relation: $KV \approx 25 k_B T_B$. However, for BFO the mean size of weak ferromagnetic domains is $d \approx 30$ nm [7] and the anisotropy constant equals $K = 6 \cdot 10^4$ J/m$^3$ [3] so this formula gives an unrealistic blocking temperature of $T_B \approx 5000$ K. Moreover, in the case of BFO sample investigated, the field dependence of $T_B(H)$ temperature cannot be fitted with usual relations describing the AT-line in the form of $T_B \sim H^{2/3}$, predicted for the SP or SG models of noninteracting spin system [24] and $T_B \sim H^{1/2}$ in the Néel model with ferromagnetic interactions [25]. Instead, the relation of $T_B$ temperature vs. $H$ is linear when presented in a semilogarithmic plot $\ln T_B - H$ (see inset to Fig. 5). The linear dependence of $\ln T_B$ on $H$ means that the relation between the temperature of bifurcation $T_B$ and the applied magnetic field is:

$$H = H_0 \exp[-bT_B(H)] \quad (2)$$

Here $H_0$ is the magnetic field value that suppresses the bifurcation temperature $T_B$ to absolute zero and b is a phenomenological parameter. It turns out that the relation (2) is similar to the formula for the energy density of ferromagnetic domain wall pinning: $E = E_0 \exp(-bT)$ [26, 27]. Therefore, we can deduce that $T_B$ is the temperature of pinning or blocking of the motion of FM or SFM domain walls rather than the SP, SG or the SSG freezing temperature $T_f$.

In the model of weak ferromagnetic domains, above $T_B$ the thermal energy exceeds the energy of domain wall pinning, which implicates free motion of domain walls and/or the domain growth. In this regime ZFC and FC $1/\chi$ curves are identical and magnetization is reversible. Below $T_B$ the thermal energy is not enough to overcome the pinning energy barrier. The domain walls are pinned and their conformation depends on the applied magnetic field and on history of the sample. Namely, when the system is field cooled it assumes the domain configuration close to the equilibrium single domain state, whereas when the system is zero field cooled it exhibits a spontaneous polydomain state. Upon ZFC, the magnetization is lower than upon FC but it starts to increase as temperature increases because the thermally activated domain wall motion becomes more rapid. The volume of the energetically favorable domains increases at the expense of other domains and the system tends to a monodomain state as for FC case. The minimum in the ZFC inverse of susceptibility (maximum in ZFC susceptibility) appears at the mean blocking temperature $T_{B,mean}$, whereas the temperature $T_B$ of bifurcation between ZFC and FC curves corresponds to the maximum blocking temperature above which the thermal energy exceeds the energy of pinning and the motion of domain walls is no longer hindered. A large difference between $T_B$ and $T_{B,mean}$ usually happens in the ferromagnetic systems with a wide distribution of domain sizes and pinning energies. For our BFO sample studied $T_B - T_{B,mean} \approx 6$ K which confirms that the distribution of domain sizes is quite narrow [13].

The blocking of domain wall motion influences also the coercivity $H_c$ and remanence $M_r$ of the hysteresis loops, which is presented in Fig. 6. In high temperature range where the domain wall pinning is ineffective, both the coercivity and remanence are negligible. However, these quantities start to increase rapidly below 100 K, a value comparable to the temperature of blocking $T_B$ determined using the criterion of ZFC and FC magnetization bifurcation (see Fig. 5). The maximum value of remanence $M_r$ for BFO sample reported here, measured at lowest temperature available i.e. 2 K was 6.1 Am$^2$/kg. The coercivity determined at the same temperature was about 0.09 T.

Besides the irreversible properties, the systems like disordered FM phase, SG, SP and SFM phases exhibit also strong time evolution of magnetic properties [28, 29]. The results of measurements of magnetic moment relaxation on time in our BFO sample are shown in Fig. 7. To better visualize the magnetic moment changes in time, all data in Fig. 7 are normalized so that the initial value of magnetic moment $m(\tau)$ for $\tau = 1$ is always equal to unity. The variable $\tau = t/t_0$ is the time normalized with respect to the dead time $t_0 \approx 10$ s needed to stabilize magnetic field and to complete the first measurement.

The time evolution of the magnetic moment $m(\tau)$ was measured during three following experiments, which allows the study of different mechanisms of magnetic relaxation:

a) ZFC to selected low temperature followed by rapid switching on of the magnetic field
b) FC to a given low temperature followed by switch off of the magnetic field
c) FC in the magnetic field to the lowest temperature available (2 K) followed by temperature increase to a selected temperature at which the measurement was next performed

The magnetic field in all experiments was 0.1 T, which corresponds to the blocking temperature $T_B$ of about 70 K.

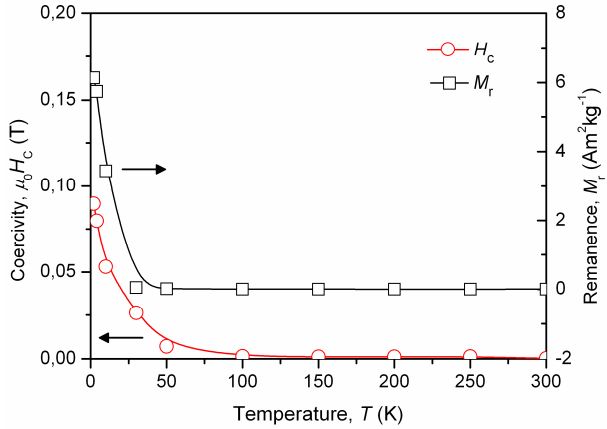

Figure 6. The dependence of coercivity $H_c$ and remanence $M_r$ of BFO sample on temperature. The solid lines are guides for eyes, only.

In the case "$a$", a time increase in the magnetic moment $m(\tau)$ is observed because of the progress in formation of domain aligned state due to domain reorientations, growth and/or domain wall motion in response to the applied magnetic field (Fig. 7a). When this process is completed the magnetic moment attains maximum value and stabilizes. However, after a long time ($\tau \approx 10^3$ corresponding to $t \approx 10^4$ s) a small decrease in the magnetic moment appears. The time evolution of the magnetic moment due to domain reorientation, until it reaches a maximum, can be described in terms of a modified power law relaxation:

$$m(\tau) = m_0 + m_R \tau^{n-1} \qquad (3)$$

where the parameters $m_0$, $m_R$ define the initial magnetization and the relaxing part of the magnetic moment and $n$ is the fitting co-efficient. The best fit of the model (3) to the experimental data was obtained for $m_0=0.57$, $m_R=0.43$ and $n=1.033$. Equally good fit can be also obtained by means of the saturating stretched exponential law used by Chen et al. [30] to describe the SFM domain systems:

$$m(\tau) = m_0 + m_R\left[1 - \exp(-\tau/\tau_0)^\beta\right] \qquad (4)$$

where $\tau_0$ and $\beta$ are fitting parameters. For the best fit (not shown for clarity in Fig. 7a) for 10 K they are as follows: $m_0=0.959$, $m_R=0.142$, $\tau_0=13$ and $\beta=0.42$. However, model (4) besides SFM phase also correctly describes relaxation in the SG phase [20, 31] and in spin cluster glass [32] and therefore gives no indication as to which phase actually occurs in the sample investigated.

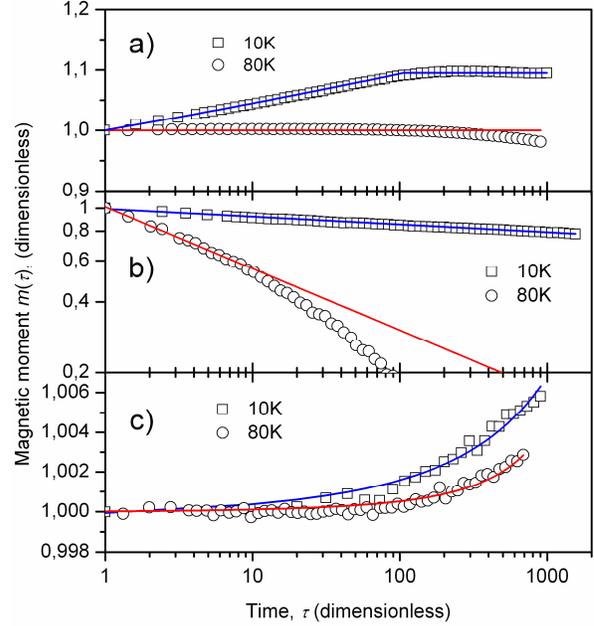

Figure 7. Time evolution of the normalized magnetic moment observed in the experiments "$a$", "$b$" and "$c$" (panels a, b and c, respectively).

Magnetic measurements performed at 80 K indicate no relaxation except a small decay of the magnetic moment for very long time, a case which will be discussed latter. At 80 K the magnetic moment immediately attains a constant value because the domain walls are not pinned above the bifurcation temperature $T_B$ and thus the process of domain ordering and domain wall motion is instantaneous. After this process of very rapid saturation of magnetic moment, the moment does not change during three decades of time (corresponding to two decades of normalized time $\tau$). Next, for a very long time another mechanism of relaxation appears, which leads to a decrease in the moment. Similar, nonmonotonic relaxations have also been observed for SFM systems [33] and attributed to intradomain thermally activated relaxation.

The panel "$b$" in Fig. 7 shows a decrease in the magnetic moment with time after FC and switching off of the magnetic field according to procedure "$b$". When the data are represented in $\log m(\tau)$-$\log \tau$ plot, they ordinate linearly, except the measurements performed at 80 K. Therefore the relaxation proceeds according to the power-law relation:

$$m(\tau) = m_0 + m_R \tau^{1-n} \qquad (5)$$

where $m(\tau)$ denotes the normalized moment, $m_0$ is the remnant magnetic moment and $m_R$ is the initial value of the relaxing part of the magnetic moment. The best fit is obtained for $m_0 \approx 0$, $m_R=1$ and $n=1.033$. The last co-efficient is identical to the $n$ parameter obtained from the fit of model (3) to the data recorded in experiment "*a*". The power-law dependence of magnetic moment on time equation (5), similar to that observed in the BFO sample studied, has been predicted theoretically for an assembly of single ferromagnetic nanodomains with dipolar interactions [34] and also found experimentally in SFM systems [30]. The power-low relaxation it is also a strong argument for SFM behaviour [8]. The value of the co-efficient $n>1$, indicates weak SFM state with ferromagnetic domains, which can be separated by ferromagnetic walls.

The values of $n$ co-efficient determined in experiments "*a*" and "*b*", are almost identical and this also supports the interpretation of the results obtained in terms of relaxation driven by domain reorientations and domain wall motion. The negligible value of $m_0 \approx 0$ determined from the best fits to the experimental data by means of model (5), corresponds to the systems exhibiting low concentration of ferromagnetic domains and no long range FM ordering. This is consistent with the conclusions educed from the Arrot construction, Fig. 4, which shows no net ferromagnetic order and the absence of interdomain ferromagnetic interactions even at lowest temperatures available in our experiment.

The mechanism of magnetic relaxation in experiment "*b*" can be understood in terms of temporal decomposition of initial almost single domain state, obtained after FC, to a polydomain state which appears immediately after the field is suppressed. This sample exhibits no remnant moment $m_0$ after long time measurements, which confirms that BFO is a weak ferromagnet with negligible inter domain interactions. At the 80 K, *i.e.* above the bifurcation temperature $T_B$, the pinning of magnetic domains vanishes and the relaxation cannot be described by means of the power-law behaviour (4) because it proceeds according to different mechanism like relaxation in SG phase.

The results of measurements after FC in a magnetic field of 0.1 T obtained in experiment "*c*" are shown in Fig. 7c. Both at 10 K and 80 K, the magnetic moment only slightly increases with time. The total change in the moment is below 1% of its initial value and becomes noticeable for long time measurements only. Therefore this slow relaxation cannot be explained by a relatively fast domain wall motion and/or domain reorientations and has probably the same origin as the long-time processes observed in experiment "*a*". It turns out that the data in Fig. 7c can be well fitted using a modification of the stretched exponential law equation (6):

$$m(\tau) = m_0 + m_R \exp(\tau/\tau_0)^\beta \qquad (6)$$

where the parameters of the best fit at 10 K (or 80 K) are: $m_0=0.9989$ (0.9998), $m_R=6.7 \cdot 10^{-4}$ ($1.5 \cdot 10^{-4}$), $\tau_0=27$ (identical at 10 K and 80 K), $\beta=0.25$ (0.27). The stretched exponential model usually well describes intra domain relaxation due to SG phase or spin cluster phase. This allows us to suppose that in BFO compound actually there are two processes of magnetization relaxation: one stronger that occurs at low temperatures due to domain reorientation, growth and/or pinning and motion of domain walls and another more subtle which dominates long term intradomain relaxation caused by SG phase. This last conclusion is consistent with the point of view of Singh et. al [10, 11] about a possible presence of SG phase in BFO compound.

The effects of aging and rejuvenation are shown in Fig. 8. To study these effects, the BFO sample was first FC in the applied field of 0.1 T to 50 K. Next the cooling was interrupted and the relaxation of magnetic moment was measured during the period $t_w=10^4$ s (data represented by open squares in Fig. 8). After that the cooling was resumed until low temperature was reached and in next step the magnetization was measured when heating the sample (filled circles in Fig. 8). The decay of magnetic moment, during the process of aging is very small, below 1% of the total signal (see the inset to Fig. 8). Thus the magnitude of this decay is comparable to the long-time variations of the magnetic moment observed in experiment "*c*" and probably driven by the same mechanisms of intradomain relaxation caused by SG phase or spin cluster glass. However, contrary to the experiment "*c*") where an increase in the magnetic moment has been observed, this time the moment decreases. It seems that there is a simple correlation between the magnetic moment evolution and the history of the sample. Namely, if the initial state of the sample exhibits magnetization (as in experiment "*c*" in which temperature was increased) higher than that in the state in which heating was interrupted and aging is studied, than the magnetic moment increases with time. Conversely, if the initial state exhibits a magnetic moment lower than that in the state in which relaxation is measured (as in the latter case in which temperature was decreased) than temporary decay of the magnetic moment is observed.

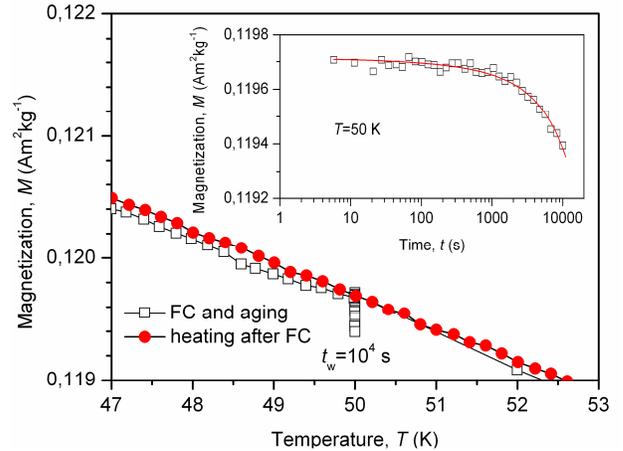

Figure 8. Aging and rejuvenation of BFO sample (main panel). Inset shows decay of the magnetic moment.

## IV. CONCLUSIONS

The results of present studies of ZFC and FC magnetization dependence on temperature, magnetization dependence of field $M(H)$ and relaxation processes after various procedures applied seem to be well explained in terms of two different mechanisms of magnetic relaxation in bismuth ferrite. The first mechanism of rapid magnetic relaxation occurs at low temperatures and is related to domain growth and/or domain wall pinning and motion in the system of weak FM or SFM domains. This point of view is supported by a field dependence of the temperature $T_B(H)$ of bifurcation between ZFC and FC magnetization curves which can be understood in terms of the model of effective energy of domain wall pinning. The power-law dependence of the magnetic moment on time is another strong argument for the relaxation to take place due to FM domains or SFM phase, whereas a negligible value of the remnant magnetic moment $m_0$ indicates a low concentration of domains. The second mechanism is long-term weak relaxation of the magnetic moment which appears in SG phase as suggested earlier by Singh et al. [10] or spin cluster phase. In this case the time dependence of magnetic moment follows the usual stretched-exponential dependence on time observed for the SG phase. This mechanism is dependent on the history of the sample in such a way that magnetization gradually relaxes towards its value exhibited in previous state and is responsible also for a nonmonotonic dependence of the magnetic moment on time.

## V. ACKNOWLEDGEMENTS

This project has been supported by National Science Centre (project No. N N507 229040) and partially by COST Action MP0904.

## REFERENCES

[1] Smolenskii G, Yudin V, Sher E, Stolypin YE: Antiferromagnetic properties of same perovskites. Sov Phys JETP 1963;16:622-624

[2] Fiebig M: Revival of magnetoelectric effect. J Phys D: Appl Phys. 2005;38:R123-R152

[3] Kadomtseva AM, Popov YuF, Pyatakov AP, Vorobev GP, Zvezdin AK, Viehland D: Phase transitions in multiferroic BiFeO$_3$ crystals, thin-layers and ceramics: enduring potential for a single phase, room-temperature magnetoelectric "holy grail". Phase Trans. 2006;79:1019-1042

[4] Lebeugle D, Colson D, Forget A, Viret M, Boville P, Marucco JM, Fusil S: Room temperature coexistence of large electric polarization and magnetic order in BiFeO$_3$ single crystals. Phys Rev B 2007;76:024116-1-8

[5] Catalan G, Scott JF: Physics and applications of bismuth ferrite. Adv Mat. 2009;21:2463-2485

[6] Sosnowska I, Neumaier TP, Steichele E: Spiral magnetic ordering in bismuth ferrite. J Phys C: Solid State Phys. 1982;15:4835-4846

[7] Ramazanoglu M, Laver M, Ratcliff W, Watson SM, Chen WC, Jackson A et al.: Local weak ferromagnetic in single-crystalline ferroelectric BiFeO$_3$. Phys Rev Lett. 2011;107:207206 1-5

[8] Petracic O, Chen X, Bedanta S, Kleemann W, Sahoo S, Cardoso S, Freitas PP: Collective states of interacting ferromagnetic nanoparticles. J Magn Magn Mater. 2006;300:192-197

[9] Gaareva ZA, Zvezdin AK: Pinning of magnetic domain walls in multiferroics. Europhys Lett. 2010;91:47006-1-3

[10] Singh MK, Prellier W, Singh MP, Katiyar RS, Scott JF: Spin-glass transition in singlecrystal BiFeO$_3$. Phys Rev B. 2008;77:144403-1-5

[11] Singh MK, Katiyar RS, Prellier W, Scott JF: The Almeida-Thouless line in BiFeO$_3$: is bismuth ferrite a mean field spin glass?. J Phys Condens Matter. 2008;21:042202-1-5

[12] Catalan G, Seidel J, Ramesh R, Scott JF: Domain wall nanoelectronics. Rev Mod Phys. 2012;84:119-156

[13] Shen TD, Schwarz RB, Thompson JD: Paramagnetism, superparamagnetism, and spinglass behavior in bulk amorphous Pd-Ni-Fe-P alloys. J Appl Phys. 1999;85:4110-4119

[14] Arrot A: Criterion for ferromagnetism from observations of magnetic isotherms. Phys Rev. 1957;108:1394-1396

[15] Kaul SN: Static critical phenomena in ferromagnets with quenched disorder. J Magn Magn Mater. 1985;53:5-53

[16] Banerjee BK: On a generalised approach to first and second order magnetic transitions. Phys Lett. 1964;12:16-17

[17] Ederer C, Spaldin NA: Weak ferromagnetism and magnetoelectric coupling in bismuth ferrite. Phys Rev B. 2005;71:060401-1-4(R)

[18] Park TJ, Papaefthymiou GC, Viescas AJ, Moodenbaugh AR, Wong SS: Size-dependent magnetic properties of single-crystalline multiferroic BiFeO$_3$ nanoparticles. Nano Letters. 2006;7:766-772

[19] Sasaki M, Jönsson PE, Takayama H, Mamiya H: Aging and memory effects in superparamagnets and superspin glasses. Phys Rev B. 2005;71:104405-1-9

[20] Suzuki M, Fullem SI, Suzuki IS, Wang L, Zhong CJ: Observation of superspin-glass behavior in Fe$_3$O$_4$ nanoparticles. Phys Rev B. 2007;79:024418-1-7

[21] Sahoo S, Petracic O, Kleemann W, Nordblad P, Cardoso S, Freitas PP: Aging and memory in a superspin glass. Phys Rev B. 2003;67:214422-1-5

[22] Du J, Zhang B, Zheng RK, Zhang XX: Memory effect and spin-glass-like behavior in Co-Ag granular films. Phys Rev B. 2007;75:014415-1-7

[23] Bitoh T, Ohba K, Takamatsu M, Shirane T, Chikazawa S: Field-cooled and zero-fieldcooled magnetization of superparamagnetic fine particles in Cu$_{97}$Co$_3$ alloy: comparison with spin-glass Au$_{96}$Fe$_4$ alloy. J Phys Soc Jpn. 1995;64:1305-1310

[24] de Almeida RL, Thouless DJ: Stability of the Sherrington-Kirkpatrick solution of a spin glass model. J Phys A: Math Gen. 1978;11:983-990

[25] Wohlfarth EP: The magnetic field dependence of the susceptibility peak of some spin glass materials. J Phys F: Metal Phys. 1980;10:L241-L246

[26] Vértesy G, Tomáš I: Temperature dependence of the exchange parameter and domainwall properties. J Appl Phys. 2003;93:4040-4044

[27] Kersten M: Über die Bedeutung der Versetzungsdichte für die Theorie der Koerzitivkraft rekristallisierer Werkstoffe. Z Angew Phys. 1956;8:496-502

[28] Vincent E, Dupuis V, Alba M, Hammann J: Aging phenomena in spin-glass and ferromagnetic phases: domain growth and wall dynamics. Europhys Lett. 2000;50:674-680

[29] Maignan A, Sundaresan A, Varadaraju UV, Raveau B: Magnetization relaxation and aging in spin-glass (La,Y)$_{1-x}$Ca$_x$MnO$_3$ (x=0.25, 0.3 and 0.5) perovskite. J Magn Magn Mater. 1998;184:83-88

[30] Chen X, Kleemann W, Petracic O, Sichelschmidt O, Cardoso S, Freitas PP: Relaxation and aging of superferromagnetic domain state. Phys Rev B. 2003;68:054433-1-5

[31] Jaeger C, Bihler C, Vallaitis T, Goennenwein STB, Opel M, Gross R, Brandt MS: Spinglass like behavior of Ge:Mn. Phys Rev B. 2006;74:045330-1-10

[32] Freitas RS, Ghivelder L, Damay F, Dias F, Cohen LF: Magnetic relaxation phenomena and cluster glass properties of La$_{0.7-x}$Y$_x$Ca$_{0.3}$MnO$_3$ manganites. Phys Rev B. 2001;64:144404-1-6

[33] de Julián C, Emura M, Cebollada F, González JM: Interactions and magnetic viscosity: Nonmonotonic time variation of the magnetization during relaxation at constant demagnetizing field. Appl Phys Lett. 1996;69:4251-4253

[34] Ulrich M, García-Otero J, Rivas J, Bunde A: Slow relaxation in ferromagnetic nanoparticles: Indication of spin-glass behaviour. Phys Rev B. 2003;67:024416-1-4